\documentclass[9pt,twocolumn,twoside]{osajnl}
\journal{ol} 

\setboolean{shortarticle}{true}

\usepackage{lineno}

\title{Optical pumping effects on the Rydberg EIT spectrum}

\author[1]{Hsuan-Jui Su}
\author[1]{Jia-You Liou}
\author[1]{I-Chun Lin}
\author[1,2,*]{Yi-Hsin Chen}
\affil[1]{Department of Physics, National Sun Yat-Sen University, Kaohsiung 80424, Taiwan}
\affil[2]{Center for Quantum Technology, Hsinchu 30013, Taiwan}

\affil[*]{Corresponding author: yihsin.chen@mail.nsysu.edu.tw}

\begin{abstract}
We provide a universal discussion of the interplay between Rydberg-state electromagnetically induced transparency (EIT) and optical pumping (OP) in a thermal $^{87}$Rb medium. By pumping the population to one single Hyperfine/Zeeman state, we can enhance the interaction strength and, in principle, amplify the EIT peak. According to our measurements, the EIT peak height can be improved by a factor of two or reduced by one order of magnitude, and linewidth was slightly narrowed by the pumping effect. Similar behavior is also seen by increasing the optical density (OD) of the medium. The EIT feature is predicted quantitatively using a Doppler-free non-perturbation numerical calculation. With and without the optical pumping field, the EIT peak heights collapse onto the same theoretical curve, showing that OP and varying OD have the same effect. In both simulations and measurements, Rydberg EIT enhancement through OP is dependent on the intensity of the probe field and the OD. Our work clarifies the underlying mechanisms of optical pumping and advances Rydberg-atom research. 
\end{abstract}

\setboolean{displaycopyright}{true}

\begin{document}

\maketitle

\newcommand{\FigOne}{
\begin{figure}[t]
\centering
\includegraphics[width=1 \linewidth]{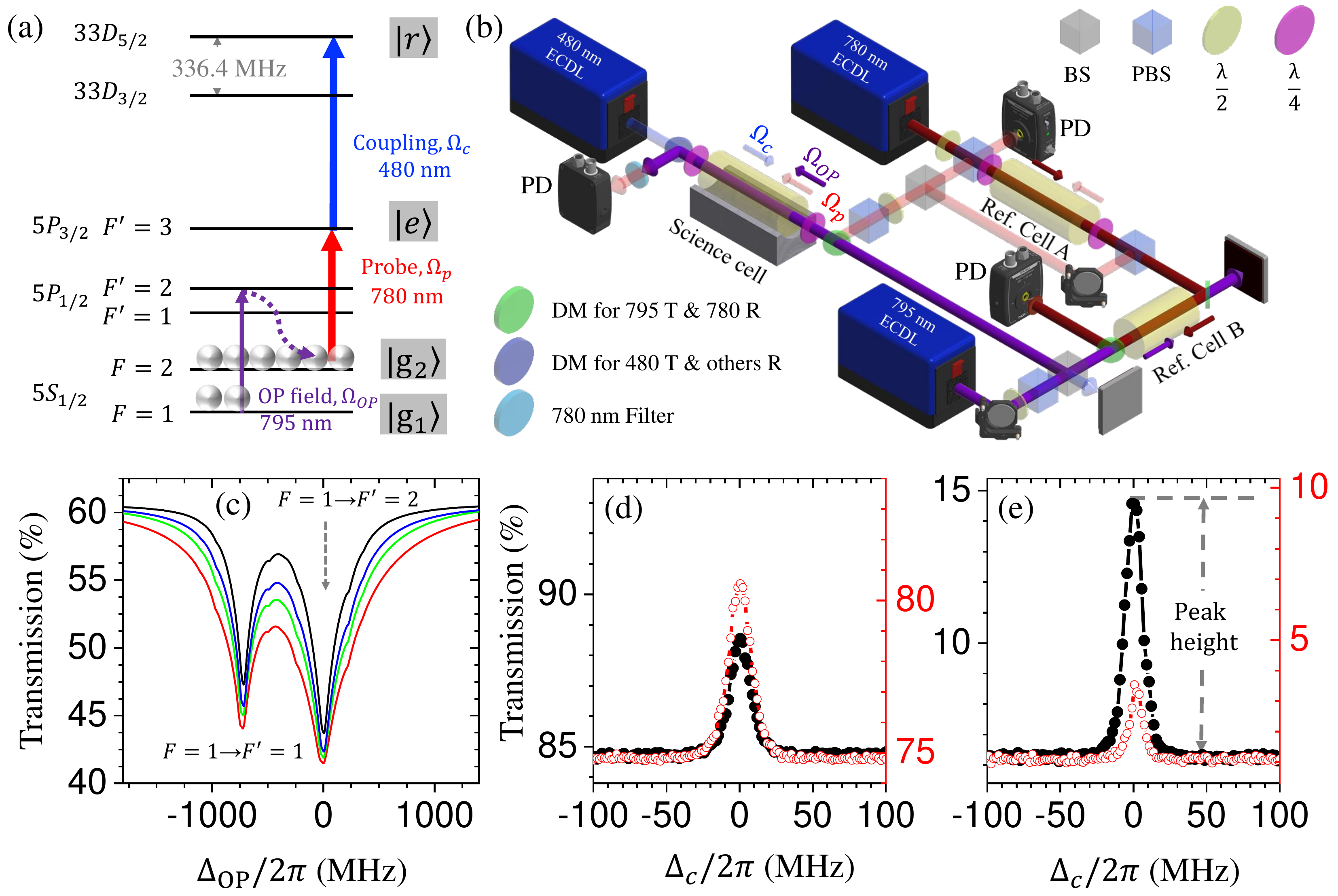}
\caption{(a) Energy level diagram for $^{87}$Rb atom involving Rydberg EIT and optical pumping transitions. A 795-nm laser ($\Omega_{OP}$) pumps the population from $|g_1\rangle$ to $|g_2\rangle$ state. A 780-nm probe field ($\Omega_p$) and a 480-nm coupling field ($\Omega_c$) form a $\Xi$-type EIT configuration. (b) Schematic of the experimental apparatus. The probe field was frequency locked via a standard saturation spectroscopy in Ref. Cell A. The pumping and probe fields were overlapped in Ref. Cell B, and the pumping field was frequency locked via the probe field transmission to the transition of $|g_1\rangle$ to $|5P_{1/2}, F'=2\rangle$ as shown in (c). With the intensity, $I_{OP}$, of 8.5 W/cm$^2$, the optical density (OD) was enhanced by 1.4~$\sim$~1.6 folds. The applied intensities shown in green, blue, and black lines were 2, 4, and 8 times weaker than the strongest one. The probe, coupling, and pumping fields were overlapped in the Science cell to perform EIT measurements. (d) and (e) were the EIT spectra with scanned coupling field across the transition of $|e\rangle$ to $|r\rangle$ with (red circles and lines) and without (black ones) the pumping field under vapor temperatures of $27^\circ$C in (d) and $63^\circ$C in (e). These measurements were performed with a probe intensity of 27 mW/cm$^2$. The spectra show that the EIT peak heights, defined as the transmission difference between the peak and baseline, can be enhanced or reduced by the optical pumping effect. DM: dichroic mirror; $\lambda/2$ and $\lambda/4$: half and quarter wave plates; (P)BS: (polarizing) cubed beam splitters. PD: photodetector.  
}
\label{fig:scheme}
\end{figure}
}
\newcommand{\FigTwo}{
\begin{figure}[t]
\centering
\includegraphics[width=0.98 \linewidth]{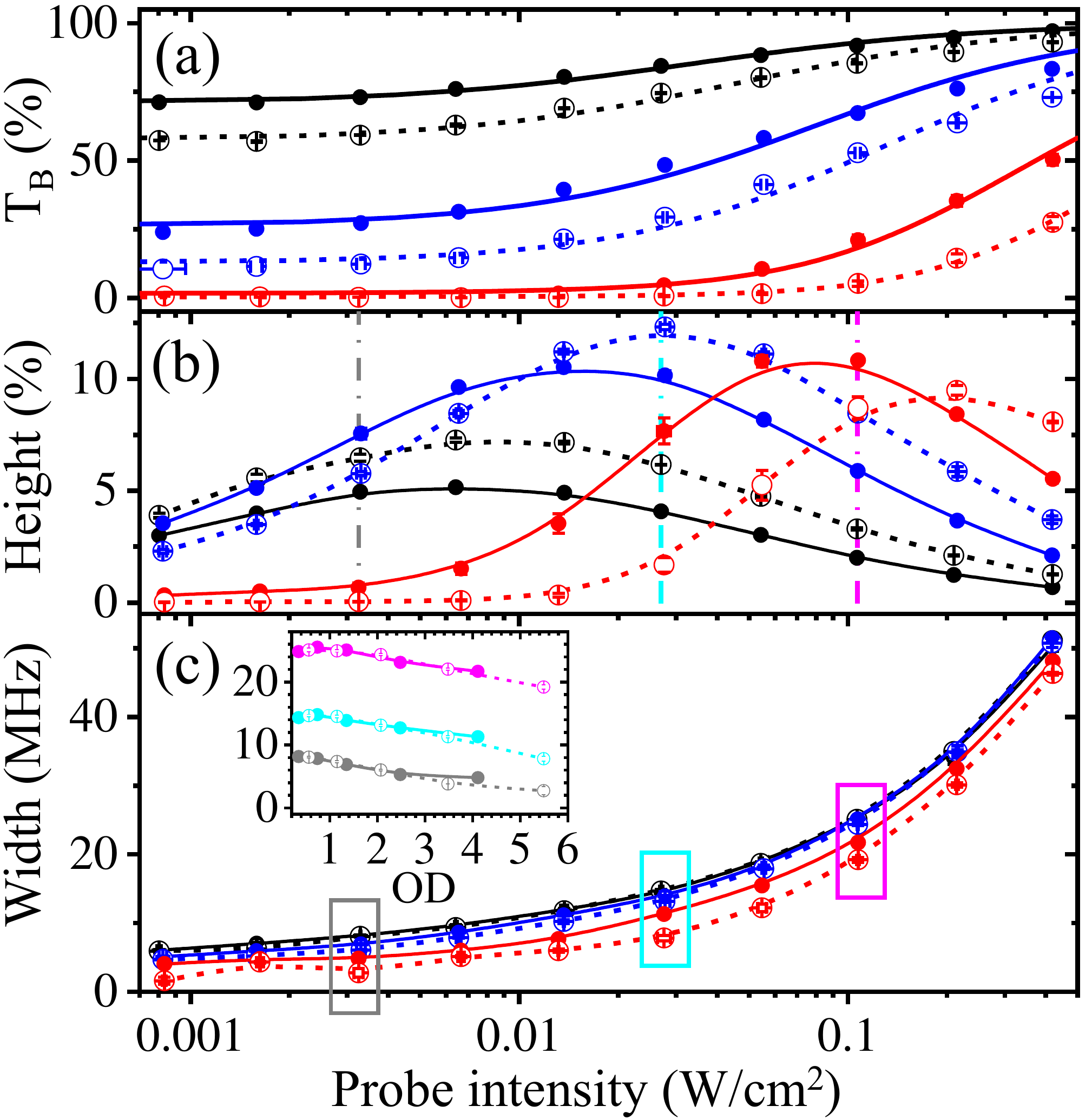}
\caption{Baseline transmission of the probe field in (a), EIT peak height in (b), and EIT linewidth in (c) with (open circles) and without (solid circles) the pumping field. Data were taken at vapor temperatures of $27^\circ$C (black circles), $48^\circ$C (blue circles), and $63^\circ$C (red circles). The solid lines in (a) are the fits from the power broadening expression by applying Eq.~(\ref{eq:broadening}). The optical pumping effect can enhance or reduce the EIT peak height, depending on the intensity of the probe field and the OD of the medium. We further extract the data with three different probe intensities (indicated by the dashed-dotted lines) and show then in Fig.~\ref{fig:OD}(b). (c) The linewidth was dominated by the probe intensity. With fixed laser intensities, the linewidth monotonously decreased with increasing OD either by heating the vapor temperature or applying an optical pumping field. The inset data are extracted with probe intensities of 3.3, 27, and 110 mW/cm$^2$ in different vapor temperatures. 
}
\label{fig:intensity}
\end{figure}
}

\newcommand{\FigThree}{
\begin{figure}[t]
\centering
\includegraphics[width=0.98 \linewidth]{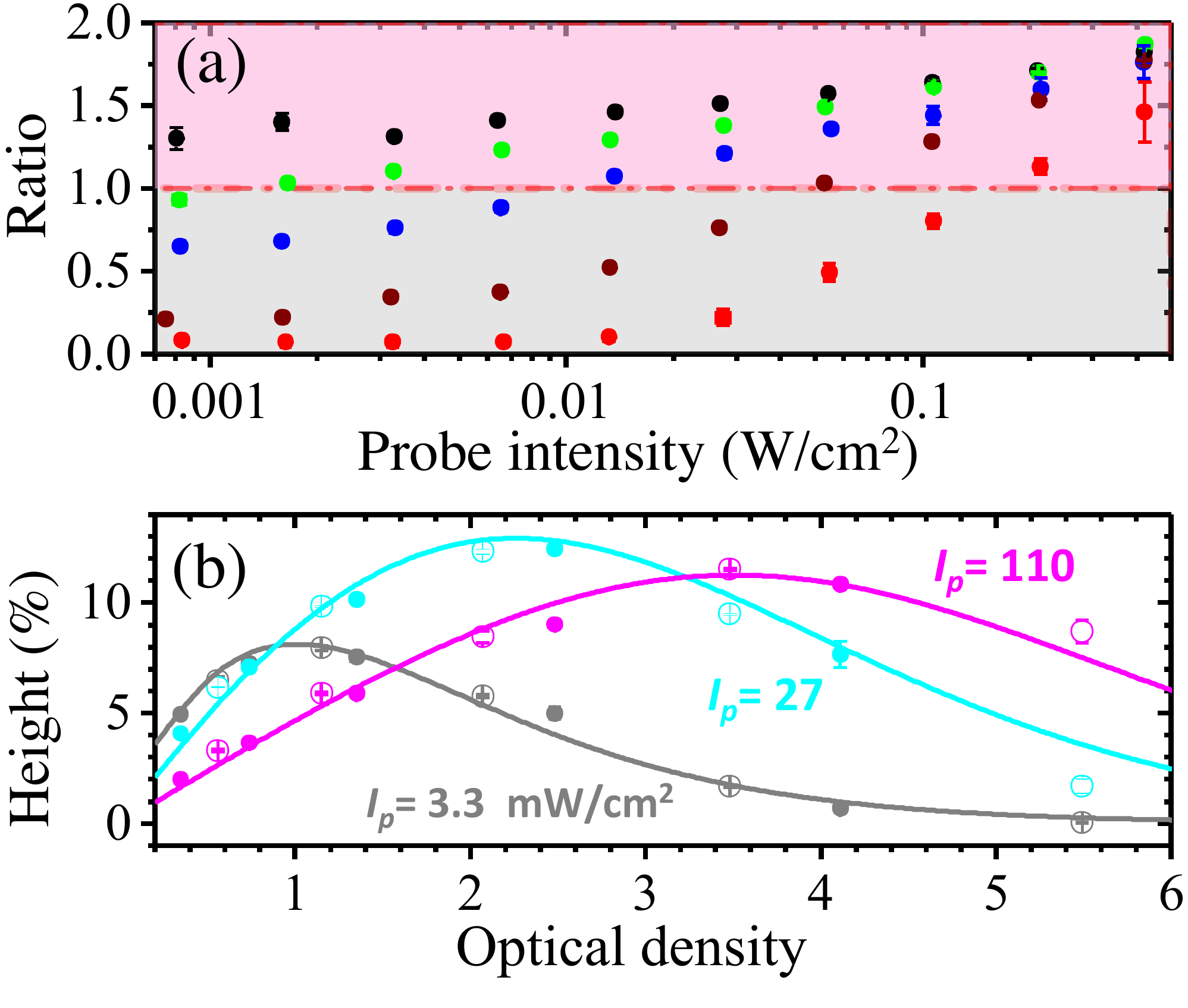}
\caption{(a) The peak height ratios of the EIT with and without population repumping in five different temperatures from 27 to 63$^\circ$C. The ODs were 0.35 (black circles), 0.74 (green), 1.4 (blue), 2.5 (brown), and 4.1 (red) before optical pumping. (b) EIT peak heights with (open circles) and without (solid circles) OP are extracted from Fig.~\ref{fig:intensity} under three probe intensities. The solid lines are the simulation by solving the OBEs and MSE in a Doppler-free non-perturbation model~\cite{Su2022}. The solid and open circles follow the same curvatures, which shows that the optical pumping and heating of the atomic vapor play the same role in the EIT feature.
}
\label{fig:OD}
\end{figure}
}

\newcommand{\FigFour}{
\begin{figure}[t]
\centering
\includegraphics[width=1 \linewidth]{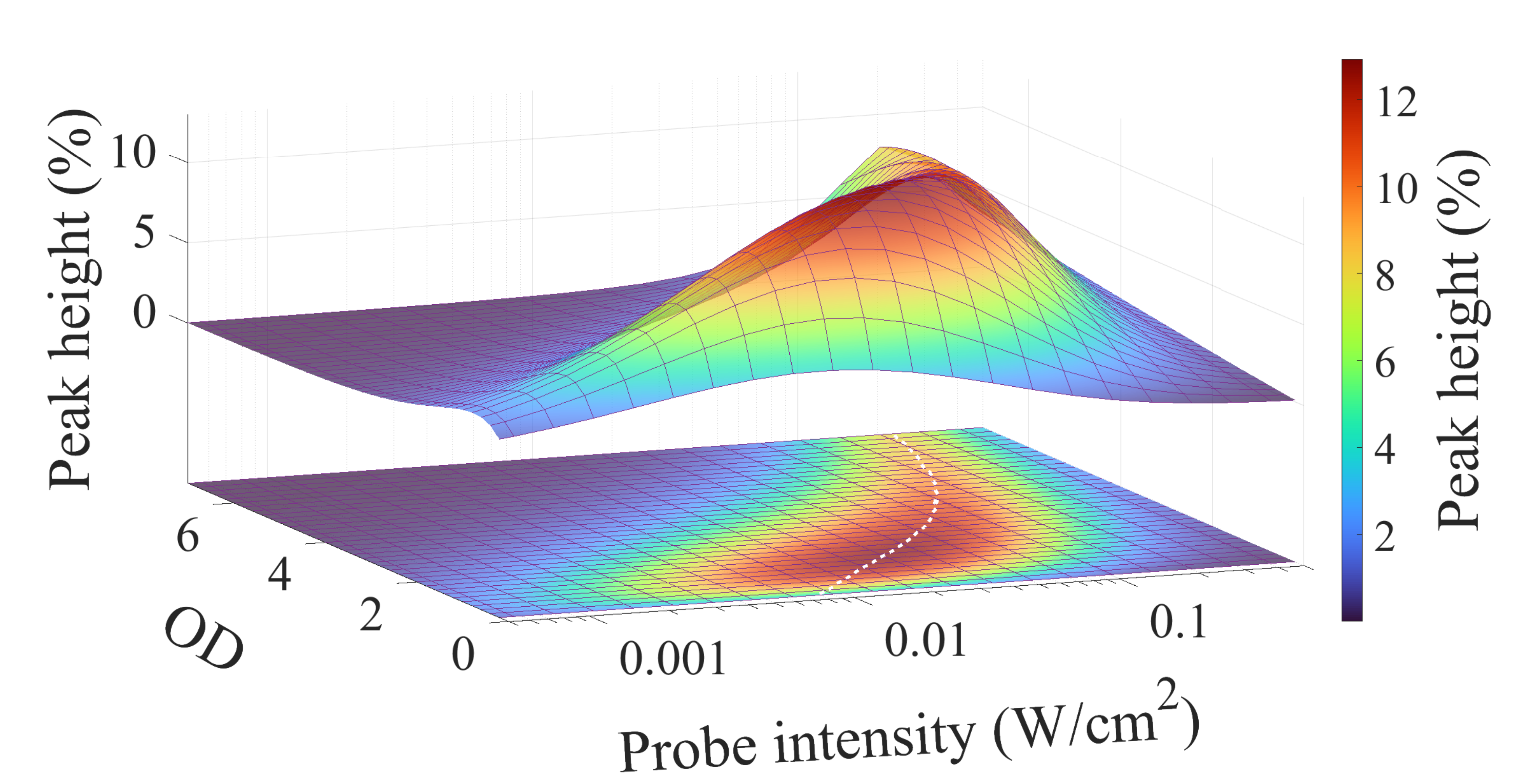}
\caption{Simulation results of EIT peak heights with a given set of OD and probe intensity. For a fixed OD ($I_p$), we observe a universal rising and falling behavior when $I_p$ (OD) increases. Besides, the optimal OD ($I_p$) to reach the maximum EIT peak height becomes stronger with a higher $I_p$ (OD). The white dashed line indicates the optimal $I_p$ for each OD. 
If the experimental condition lies below the line, increasing OD can enhance the peak height with a fixed probe intensity. 
}
\label{fig:ODinensity}
\end{figure}
}

Rydberg-atom-based spectroscopy is a technique to detect the external field from DC to THz sensitively or to measure the interactions between atoms~\cite{Sedlacek2012,Fan2015,Jiao2016,Vogt2018,Jau2020,Robinson2021,Meyer2021}. 
A narrow and high contrast electromagnetically-induced-transparency (EIT) feature is helpful to probe the energy level shifts due to the external field or the atomic interactions~\cite{Zhao2009,Kumar2017,Simons2018,Jia2020,Fancher2021}.
Combined with an enhanced EIT amplitude, the researchers in Ref.~\cite{Holloway2021} have improved the sensitivity of Rydberg electrometry by using the Autler-Townes (AT) splitting and a radio frequency local oscillator field. 
The EIT amplitude can be optimized by applying a proper probe intensity (which drives the transition of the ground and intermediate states) and a stronger coupling field (which couples that of the intermediate and Rydberg states)~\cite{Wu2017,Su2022}. By modulating the probe field frequency, the EIT linewidth can be reduced by a factor of two, and the accuracy of the absolute Rydberg transition frequencies can be improved~\cite{Silpa2022}. Therefore, a high contrast EIT spectrum with enhanced peak height and reduced linewidth can be achieved, making relevant studies useful in developing Rydberg-based electrometry.

The laser polarization configurations of EIT for Rydberg S-state and D-state orbitals have been systematically studied~\cite{Wu2017}. The EIT peak height can be improved through the optimization of laser intensities and optical densities by controlling the vapor temperature~\cite{Su2022}. Besides, the optical pumping by the laser fields can vary the EIT structure because of the population redistribution among Hyperfine/Zeeman states~\cite{He2013,Moon2008,Yang2010,Zhang2018}. The interplay between optical pumping among the Zeeman sublevels and Rydberg EIT under the presence of a magnetic field exhibits rich sets of EIT peaks~\cite{Zhang2018}. In the investigations of double-resonance optical pumping (DROP)~\cite{Moon2004,Moon2008}, the population in the ground state, e.g., $|g_2\rangle$ in Fig.~\ref{fig:scheme}(a), may be depleted since atoms are excited and then optically pumped into another ground state, e.g., $|g_1\rangle$, through the intermediate and higher excited Rydberg states.
With the population transformation, the probe beam's absorption is reduced, resulting in high-resolution DROP spectroscopy, which is useful for laser frequency stabilization. As opposed to depleting ground state populations through the DROP effect, the EIT peak can also be increased by optically pumping the population to $|g_2\rangle$ state to enhance the strength of the interaction~\cite{Holloway2021}.
According to our measurements, the optical pumping can improve or reduce the EIT peak height in different probe field intensities and optical densities.
We provide a Doppler-free non-perturbation numerical calculation to predict the EIT feature quantitatively. The EIT peak heights with and without the optical pumping field collapse onto the same theoretical curve. The underlying mechanisms of optical pumping and the improvement of Rydberg-state EIT are clarified in this Letter. 
 
\FigOne

We performed the optical pumping effect of Rydberg EIT in a thermal $^{87}$Rb vapor. Figure~\ref{fig:scheme}(a) shows the energy level diagram involving EIT and optical pumping transitions, and Fig.~\ref{fig:scheme}(b) is the corresponding schematic of the experimental apparatus. 
A 795-nm laser ($\Omega_{OP}$, Toptica DL pro) pumps the population from ground state $|5S_{1/2}, F=1\rangle$ (denoted as $|g_1\rangle$) to $|5S_{1/2}, F=2\rangle$ (denoted as $|g_2\rangle$) via an intermediate state $|5P_{1/2}, F'=2\rangle$. A 780-nm probe field ($\Omega_p$, Toptica DL pro) and a 480-nm coupling field ($\Omega_c$, Toptica DL pro HP 480) form a $\Xi$-type EIT configuration. All of the spectra were real-time measurements within one second via the PZT scanning of the external cavity diode laser (ECDL) of the coupling field, taken directly from the Toptica DLC pro.
The probe field was frequency locked via a reference cell (Cell A) to the transition of ground state $|g_2\rangle$ to the intermediate state $|5P_{3/2}, F'=3\rangle$ (denoted as $|e\rangle$) using the standard sub-Doppler saturation spectroscopy. 
If the population is equally distributed between two Hyperfine states initially, then the atomic number in $|g_2\rangle$ state can be doubled when applying the pumping field. 
A second reference cell (Cell B) was used to overlap the probe and pumping fields, and the pumping field was frequency locked using the probe transmission, as shown in Fig.~\ref{fig:scheme}(c).
The pumping field was swept across the transition of $|g_1\rangle$ to $|5P_{1/2}, F'=1\rangle$ and $|5P_{1/2}, F'=2\rangle$. It was clear that driving to $|F'=2\rangle$ state has a better population pumping efficiency. With the intensity of 8.5 W/cm$^2$, the optical density (OD) was improved by 1.4~$\sim$~1.6 folds. 
Despite lowering the pumping intensity, the variation in OD at resonant frequency was minimal, indicating that the applied intensity was sufficiently strong.
The probe, coupling, and pumping fields were collimated to the full width at $e^{-2}$ maximum of 0.81, 1.4, and 0.72 mm, respectively, overlapping in the Science cell (Thorlabs GC25075-RB) to perform EIT measurements.
Note that the beam size has been optimized to achieve the best EIT peak height. All three laser fields have $\sigma^+$ polarization.
We obtained the typical Rydberg-EIT spectra by scanning the coupling field frequency across the transitions of $|e\rangle$ to $|33D_{3/2}\rangle$ and $|33D_{5/2}\rangle$. The frequency difference between them is 336.4 MHz~\cite{Mack2011}, which is used to calibrate the coupling field detuning in the whole measurements. Figures \ref{fig:scheme}(d) and \ref{fig:scheme}(e) were the $|33D_{5/2}\rangle$ (denoted as $|r\rangle$) Rydberg EIT spectra with (red circles and lines) and without (black ones) the pumping field under vapor temperatures of $27^\circ$C in (d) and $63^\circ$C in (e) with fixed probe intensity of 27 mW/cm$^2$. The spectra show that the optical pumping effect can enhance or reduce the EIT peak heights. The universal behavior of the OP effect will be discussed.
\FigTwo


To systematically study the effect of optical pumping, we varied the probe field intensity and vapor temperature, and extracted the baseline transmission (the probe transmission at the coupling frequency detuned far away from the two-photon resonance), EIT peak height (the transmission difference between the peak and baseline), and EIT linewidth (full width at half maximum), as shown in Fig.~\ref{fig:intensity}. 
The EIT characteristics with and without the OP field are indicated in open and solid circles, respectively. 
The figure shows three of the five measurements taken at different vapor temperatures from $27^\circ$C to $63^\circ$C to control the atomic density as well as the optical density.
By considering the power broadening effect, the optical density (denoted as $\alpha$) is determined from the baseline transmission, $T_B$, by the expression.
\begin{equation}
T_B = \exp \left(-\alpha~\frac{\Gamma_e^2}{\Gamma_e^2+2 \Omega_p^2} \right),\\
\label{eq:broadening}
\end{equation}
where $\Gamma_e$ is the effective linewidth of the excited state and $\Omega_p$ is the Rabi frequency of the probe field. Solid and dashed lines in Fig.~\ref{fig:intensity}(a) are the best fits. ODs were 0.35 (0.56), 1.4 (2.1), and 4.1 (5.5) for black, blue, and red solid (open) circles, respectively, indicating that OP increased OD by 1.4~$\sim$~1.6 folds. 

The EIT peak height measurements gave a universal phenomenon: when we increased the probe field intensity, the peak height initially enhanced, then reached the maximum at the optimal $I_p$, and finally decreased. As discussed in Refs.~\cite{Wu2017,Su2022}, the power broadening effect slightly raises the baseline transmission and the coherent population transformation from the ground state to the Rydberg state (the DROP effect) also monotonously improves the EIT peak transmission when increasing $I_p$ ($\propto \Omega_p^2$). 
Even though the peak transmission becomes saturated at a large $I_p$, the baseline transmission can still be significantly increased.
Consequently, an increment of $I_p$ makes the baseline transmission approach to the EIT peak transmission, implying the peak height decreases to zero. For any given temperature, the peak height reaches its maximum value with the optimal $I_p$, which also became stronger as the OP field was applied. 
In low optical density conditions, e.g., OD of 0.35 shown in black circles, the optical pumping effect significantly increased the EIT peak height for all probe intensities. In contrast, the effect also reduces peak height in high OD and low probe intensity conditions, e.g., OD of 4.1 shown in red circles. 
In other words, while optically pumping the population to a single state increases interaction strength, the EIT feature is dependent on experimental conditions such as laser intensity and OD. 
As a result of optical pumping, the baseline transmission and EIT peak height are greatly influenced, whereas the linewidth is dominated by the probe field intensity, shown in Fig.~\ref{fig:intensity}(c). In the theoretical perturbation model ($\Omega_p\ll \Omega_c$)~\cite{Fleischhauer2005}, the EIT linewidth is a function of the coupling field Rabi frequency $\Omega_c$ and OD (linewidth $\propto \Omega_c^2/\sqrt{
\alpha}$). Thus, a higher OD (by increasing vapor temperature or applying OP field) resulted in a narrower linewidth, as shown in the inset.

\FigThree

Optical density can be controlled by both optical pumping and vapor temperature. 
Optical pumping accumulates the population to a single Hyperfine/Zeeman state, while a higher vapor temperature leads to more atoms transitioning into a vapor as the kinetic energy increases. Optical density is equal to $n\sigma L$, where $n$ is the atomic density, $\sigma$ is the absorption cross-section, and $L$ is the light-matter interaction length. The peak height ratios of the EIT with and without OP are shown in Fig.~\ref{fig:OD}(a), taken in five different temperatures from $27$ to $63^\circ$C. The peak height can be improved by a factor of two or reduced by one order of magnitude by the OP effect.  
We further extract the Rydberg EIT peak heights with (open circles) and without (solid circles) the OP beam in three probe intensities, shown in Fig.~\ref{fig:OD}(b). Optical densities are determined from the baseline transmission after considering the power broadening effect. The OP effect will reduce the peak height if the ODs are larger than 0.8, 2.2, and 3.4 for the probe intensities of 3.3, 27, and 110 mW/cm$^2$, respectively. The optimal OD to reach the maximum EIT peak height becomes larger with a higher $I_p$. Thus, we found that the OP effect can enhance the peak height in a sufficiently low OD regime or reduce it in a large OD regime. 

We applied a non-perturbation numerical calculation by solving optical Bloch equations (OBEs) and Maxwell-Schr$\ddot{\rm{o}}$dinger equation (MSE). Details can be found in our previous study~\cite{Su2022}. The solid lines are the corresponding fits, where we fix the effective coupling Rabi frequency,  $\Omega_c/2\pi = 0.44$ MHz, and the effective coherence dephasing rate $\gamma$ between the ground state and Rydberg state, $\gamma/2\pi = 0.75$ MHz. By comparing the theoretical EIT linewidths to the spectral measurements, $\Gamma_e/2 \pi$ is determined to be 60 MHz, which includes the selection of velocity groups participating in the interaction. 
The effective linewidth of the excited state $\Gamma_e$ is predominantly determined by the interaction time of the atom with the applied laser fields, i.e., the average time-of-flight of the atom through the laser beam. The conversion ratio of $I_p$ to the Rabi frequency $\Omega_p$ in the simulation was varied with OD from 4.1 to 2.8 $\Gamma_e$ for $I_p = 1$W/cm$^2$ because of the optical pumping effect of the probe field in different vapor temperatures~\cite{Su2022}. 
With the set parameters, all EIT features, including the baseline transmission, EIT peak height, and linewidth, can be well simulated. Thus, with modified effective parameters, the Doppler-free non-perturbation model can phenomenologically predict the maximum EIT peak height. 
We then fix the intensity conversion value as 3.4$\Gamma_e$ in the simulation of Figs.~\ref{fig:OD} and~\ref{fig:ODinensity}, without loss of generality. 
Through the systematic study by applying the OP field in different vapor temperatures, we found that all peak height measurements collapse to the same simulation curves for different probe intensities. 
Therefore, we demonstrate that increasing optical density by applying the optical pumping and heating the vapor play the same role in the EIT feature. 

\FigFour
Next, we discuss the interplay of Rydberg EIT and optical pumping, and the limitations of EIT enhancement. The EIT peak height simulation results are shown in Fig.~\ref{fig:ODinensity}, using the above-mentioned parameters with a given set of OD and probe intensity. 
For a fixed OD, the power broadening effect slightly raises the baseline transmission, but the coherent population transformation from $|g_2\rangle$ to $|r\rangle$ state also monotonously improves the EIT peak transmission as $I_p$ increases. Competition between the baseline and peak transmission leads to the universal rising and falling behavior. The optimal $I_p$ to reach the maximum EIT peak height becomes stronger with a higher OD.
Furthermore, for a given $I_p$, the results show similar behavior when the OD is varied. The optimal OD also increases as $I_p$ becomes stronger. 
The white dashed line indicates the optimal $I_p$ for each OD. 
If the experimental condition is below the line, the peak height can be enhanced using optical pumping or other methods to increase OD. 
The comprehensive study of the EIT feature will be helpful for laser frequency locking~\cite{Wang2019,Jia2020_2} and can enhance the sensitivity of detecting external fields and atomic interactions.   

In conclusion, we studied the optical pumping effect on the Rydberg EIT feature and demonstrated that the EIT peak height could be increased by two folds or reduced by one order of magnitude through OP. 
The optical density can be controlled by the atomic density as well as the vapor temperature, light-matter interaction length, and optical pumping. 
We conclude that the OD is the universal parameter that determines the optimal probe intensity for the maximum EIT peak height. A narrower EIT linewidth is caused by the pumping effect that increases OD. 
Rydberg EIT enhancement through OP is dependent on the intensity of the probe field and the OD in both measurements and simulations. By utilizing the Doppler-free non-perturbation numerical calculation, a quantitative theoretical model is employed to predict the EIT feature. The EIT peak heights with and without the optical pumping field collapse onto the same theoretical curve, indicating that the optical pumping and heating of the atomic vapor play the same role in increasing OD. 
Our work clarifies the underlying mechanism and provides a broader knowledge for the Rydberg-EIT studies. By optimizing the EIT amplitude, the sensitivity of the Rydberg-based electrometry sensing will also be increased. Therefore, our investigation advances future applications such as the study of highly excited Rydberg states dipole-dipole interaction, laser-frequency stabilization, and traceable detection of the environment electromagnetic field as a quantum sensor.

\begin{backmatter}
\bmsection{Funding} Ministry of Science and Technology, Taiwan (109-2112-M-110-008-MY3, 110-2123-M-006-001).

\bmsection{Acknowledgments} This work was supported by the Ministry of Science and Technology of Taiwan under Grant Nos. 109-2112-M-110-008-MY3, 110-2123-M-006-001, and financially supported by the Center for Quantum Technology from the Featured Areas Research Center Program within the framework of the Higher Education Sprout Project by the Ministry of Education (MOE) in Taiwan. 

\bmsection{Disclosures} The authors declare no conflicts of interest.
\bmsection{Data Availability Statement} Data underlying the results presented in this paper are not publicly available at this time but may be obtained from the authors upon reasonable request.

\end{backmatter}

\bibliography{RydbergEITRef}


\end{document}